\documentclass[conference,10pt]{IEEEtran}

\usepackage{pgfplots}
\pgfplotsset{compat=newest}
\usepackage{tikz}
\usetikzlibrary{arrows,matrix,positioning}
\usepackage[utf8]{inputenc}
\usepackage[ruled,linesnumbered,vlined]{algorithm2e}
\SetKwInput{kwInit}{\textbf{Init}}
\usepackage[hang,flushmargin]{footmisc}
\usepackage{graphicx}
\usepackage{lipsum}
\makeatletter
\newcommand{\algorithmfootnote}[2][\footnotesize]{%
  \let\old@algocf@finish\@algocf@finish% Store algorithm finish macro
  \def\@algocf@finish{\old@algocf@finish% Update finish macro to insert "footnote"
    \leavevmode\rlap{\begin{minipage}{\linewidth}
    {#1#2}
    \end{minipage}}%
  }%
}
\makeatother
\usepackage[english]{babel}
\usepackage{amsmath,amsbsy, amssymb}
\usepackage{mathdots}
\usepackage{mathtools}
\usepackage{url}

\usepackage{color}
\usepackage{cite}
\usepackage{tabularx}
\usepackage{footnote}

\usepackage{amsthm}
\newtheorem{definition}{Definition}%[section]
\newtheorem{corollary}{Corollary}%[section]
\newtheorem{theorem}{Theorem}%[section]
\newtheorem{lemma}{Lemma}%[section]
\newtheorem{remark}{Remark}

\usepackage{array}
\usepackage{multirow}
\usepackage{arydshln}

\usepackage[capitalise]{cleveref}
\usepackage{verbatim}

\DeclareMathOperator{\rank}{rank}

\DeclareMathOperator{\wt}{wt}
\DeclareMathOperator{\supp}{supp}

\newcommand{\F}{\ensuremath{\mathbb{F}}}
\newcommand{\Fq}{\ensuremath{\mathbb{F}_q}}

\newcommand{\power}[2]{\ensuremath{#1}^{q^{#2}}}

\newcommand{\ceil}[1]{\ensuremath{\left\lceil{#1}\right\rceil}}
\newcommand{\floor}[1]{\ensuremath{\left\lfloor{#1}\right\rfloor}}

\newcommand{\Mod}{\ \mathrm{mod}\ }
\newcommand{\tikznode}[2]{%
	\ifmmode%
	\tikz[remember picture,baseline=(#1.base),inner sep=0pt] \node (#1) {$#2$};
	\else
	\tikz[remember picture,baseline=(#1.base),inner sep=0pt] \node (#1) {#2};%
	\fi}
\tikzset{%
	mybox_block/.style={rectangle,rounded corners,draw=black, thick,text width=1em,minimum height=2em,minimum width=4.75em,text centered},
	[highlight/.style={rectangle,rounded corners,fill=#1!15,draw,fill opacity=0.5,thick,inner sep=0pt},
	highlight/.default=gray],
	plot1/.style = {thick,
		dotted,
		mark=+}
}
\newif\ifcomment
\commentfalse
\newcommand{\new}[1]{{\color{black} #1}}

\def\ve#1{{\mathchoice{\mbox{\boldmath$\displaystyle #1$}}%
              {\mbox{\boldmath$\textstyle #1$}}%
              {\mbox{\boldmath$\scriptstyle #1$}}%
              {\mbox{\boldmath$\scriptscriptstyle #1$}}}}

\newcommand{\set}[1]{\ensuremath{\mathcal{#1}}}

\newcommand{\0}{\ensuremath{\ve{0}}}

\newcommand{\C}{\ensuremath{\ve{C}}}

\newcommand{\E}{\ensuremath{\ve{E}}}

\newcommand{\G}{\ensuremath{\ve{G}}}
\renewcommand{\H}{\ensuremath{\ve{H}}}

\newcommand{\R}{\ensuremath{\ve{R}}}

\renewcommand{\c}{\ve{c}}
\newcommand{\e}{\ve{e}}

\newcommand{\h}{\ve{h}}

\renewcommand{\r}{\ve{r}}
\newcommand{\s}{\ve{s}}

\renewcommand{\v}{\ve{v}}

\newcommand{\cE}{\mathcal{E}}

\newcommand{\cI}{\mathcal{I}}
\newcommand{\cL}{\mathcal{L}}

\commenttrue
\IEEEoverridecommandlockouts
\usepackage[ancient]{flushend}
\begin{document}

\title{Decoding of (Interleaved) Generalized Goppa Codes}

\author{
	\IEEEauthorblockN{Hedongliang Liu$^{1,*}$, Sabine Pircher$^{1,2,*}$, Alexander Zeh$^2$, Antonia Wachter-Zeh$^{1}$}
	\IEEEauthorblockA{$^1$Institute for Communications Engineering, Technical University of Munich (TUM), Munich, Germany}
	\IEEEauthorblockA{$^2$HENSOLDT Cyber GmbH, Research \& Development, Ottobrunn, Munich, Germany}
	\IEEEauthorblockA{\texttt{ \small {\{lia.liu, antonia.wachter-zeh\}@tum.de}, {\{sabine.pircher, alexander.zeh\}@hensoldt-cyber.de}}}
	\vspace{-0.45cm}
	\thanks{* The first two authors contributed equally to this work.
		
		The work A.~Wachter-Zeh has been supported by the European Research Council (ERC)
		under the European Union’s Horizon 2020 research and innovation
		programme (grant agreement no.~801434).
		The work of H.~Liu has been supported by the DFG with a German Israeli
		Project Cooperation (DIP) under grant no. KR3517/9-1.}
}

\maketitle

\begin{abstract}
  Generalized Goppa codes are defined by a \emph{code locator} set $\cL$ of polynomials and a \emph{Goppa polynomial} $G(x)$.
  When the degree of all code locator polynomials in $\cL$ is one, generalized Goppa codes are classical Goppa codes.
  In this work, \emph{binary generalized Goppa codes} are investigated.
  First, a parity-check matrix for these codes with code locators of any degree is derived.
  A careful selection of the code locators leads to a lower bound on the minimum Hamming distance of generalized Goppa codes which improves upon previously known bounds.
  A quadratic-time decoding algorithm is presented which can decode errors up to half of the minimum distance.
  \emph{Interleaved generalized Goppa codes} are introduced and a joint decoding algorithm is presented which can decode errors beyond half the minimum distance with high probability. \new{Finally, some code parameters and how they apply to the \textit{Classic McEliece} post-quantum cryptosystem are shown.}
\end{abstract}
\section{Introduction}
Goppa codes~\cite{GVD70} are currently receiving renewed attention due to their applicability in the McEliece public-key cryptosystem \cite{McE78}, which has remained unbroken for more than 40 years. \emph{Generalized Goppa codes} (GGCs) are an extension of Goppa codes to a new class of codes which are defined by a set of \textit{code locator polynomials} and a \emph{Goppa polynomial} ~\cite{shekhunova_cyclic_1981,bezzateev1997gGC}.
A special class of binary GGCs which is perfect in the weighted Hamming metric was introduced in~\cite{BezzateevShekhunova2013Perfect} and cyclic GGCs were investigated in \cite{bezzateev2014subclass,bezzateev_cyclic_2015}.
Recent works~\cite{noskov_one_2020}, \cite{noskov_mceliece_2020} present a construction of binary GGCs with irreducible code locator polynomials of first and second degree.

The McEliece cryptosystem is believed to be secure against attacks of a capable quantum computer and the Niederreiter's~\cite{niederreiter_knapsack-type_1986} dual version of the McEliece cryptosystem is a finalist in the ongoing post-quantum NIST competition \cite{computer_security_division_round_2017} under the name \emph{Classic McEliece} \cite{ClassicMcElieceNIST}.
Wild Goppa codes~\cite{WildGC76} are shown to have a larger minimum distance than classical Goppa codes and are deployed in \emph{Wild McEliece}~\cite{bernstein_wild_2011}, which is also part of \emph{Classic McEliece}.
In~\cite{noskov_mceliece_2020}, \emph{Classic McEliece} using binary GGCs with code locator polynomials of first and second degree is proposed.
Compared to classical Goppa codes, the length of GGCs can be increased by using higher-degree code locators for a fixed field size or, vice versa, for a fixed length, GGCs require a smaller field size. In practice, performing computations over smaller field sizes reduces the complexity of calculations and might therefore lead to more efficient encryption and decryption procedures. \new{GGCs may also be used as locally correctable codes in coded distributed storage systems due to their local error-correction capability on locations of different degrees. In~\cite{BezzateevShekhunova1998Localized} their ability to decode localized errors and to correct more errors than classical Goppa codes (with certain failure probability) was investigated.}

In this work, we investigate binary GGCs. Our main contributions are: \new{in Section~\ref{sec:bgGC},} we derive a parity-check matrix for GGCs with code locators of any degree, \new{where an instance for GGCs with degree-2 code locators is presented in \cite{noskov_one_2020},\cite{noskov_mceliece_2020} without proof;}
\new{we provide a formal proof for the lower bound on the minimum Hamming distance of  GGCs, which was stated in \cite{noskov_one_2020},\cite{noskov_mceliece_2020} without proof and we show that the lower bound for GGCs with even-degree code locators is improved compared to the general lower bound.}
In Section~\ref{sec:UniDec} we provide an explicit decoding algorithm for GGCs and we \new{prove the unique decoding radius for GGCs.}
To deal with burst errors, we introduce \emph{interleaved generalized Goppa codes} in \cref{sec:IntDec}. We provide \new{an explicit} decoding algorithm \new{and derive the new maximum decoding radius for GGCs}.
Moreover, we list some code parameters of GGCs and discuss their applicability to the McEliece cryptosystem in Section~\ref{sec-McEliece}.
\section{Preliminaries}\label{sec:preliminaries}
We denote by $[a,b]$ the set of integers $\{i | a\leq i \leq b\}$ and if $a=1$, we omit it from our notation and write $[b]$. A finite field
of size $q$ is denoted by $\Fq$.
Row vectors are denoted by bold lower-case letters (e.g., $\c$) and column vectors by $\c^{\top}$. Denote $\supp(\c):=\{i|c_i\neq 0\}$.
Denote matrices by bold capital letters (e.g., $\C$) and its $i$-th row by $\c^{(i)}$. We consider the Hamming metric for weight and distance.
Sets are denoted by calligraphic letters (e.g., $\cL$) and its size is denoted by $|\cL|$.

Let $\Fq[x]$ denote a polynomial ring with coefficients in $\Fq$.
For a polynomial $f(x)$, its degree is denoted by $\deg f(x)$ and its formal derivative is denoted by $f'(x)$.
The greatest common divisor of two polynomials is denoted by $\gcd(f(x),g(x))$.
\begin{lemma}[Roots of Irreducible Polynomials~{\cite[p.~52]{lidl_niederreiter_1996}}]\label{lem:splitting_roots} Let $q$ be a prime power.
 Any irreducible
  polynomial $f(x)\in\Fq[x]$ of degree $k$ can be represented as
  \begin{align*}
    f(x)=(x-\beta)(x-\power{\beta}{})\cdots(x-\power{\beta}{k-1}),
  \end{align*}
  where $\beta\in\F_{q^k}$ and $\F_{q^k}$ is called the \emph{splitting field} of $f(x)$.
\end{lemma}
\begin{lemma}[Number of Irreducible Polynomials~{\cite[p.~225]{roth2006}}]\label{lem:numIrrePoly}
The number $\mathcal{I}_{q}(t)$ of irreducible polynomials of degree $t$ over $\Fq$ can be calculated by
  \begin{equation*}
	\mathcal{I}_q(t)=\frac{1}{t}\sum_{k | t} \mu(k)\cdot q^{\frac{t}{k}}
  \end{equation*}
  where $\mu(t)$ is the Möbius function (cf.~\cite[p.~224]{roth2006})
  \begin{equation*}
    \mu(t)=
    \begin{cases}
      1 & \textrm{if } t=1\\
      (-1)^s & \textrm{if } t \textrm{ is a product of }s \textrm{ distinct prime}\\ 
      0 & \textrm{otherwise.}
    \end{cases}
  \end{equation*}
\end{lemma}

\section{Binary Generalized Goppa Codes}\label{sec:bgGC}
\new{We introduce} a parity-check matrix of generalized Goppa codes (GGCs) with code locators of any degree and  \new{ provide a proof of} \new{the} lower bound on the minimum Hamming distance of GGCs which is stated in~\cite{noskov_one_2020,noskov_mceliece_2020} \new{without proof. We} show that with \new{even-degree} code locators, the lower bound can be slightly improved.
\begin{definition}\label{def:bgGC}
  Let $m,n,r,q$ be positive integers such that $rm\leqslant n$ and
  $q=2^m$. Given a polynomial $G(x) \in \Fq[x]$ of degree $r$ and
  a set of irreducible polynomials
  \begin{equation}
    \cL=\left\{{f_1(x)},f_2(x),\dots,{f_n(x)}\right\}
    \label{eq:gcc_L}
  \end{equation}
  with $\gcd(f_i(x),f_j(x))=1 , \forall i\neq j$, and $\gcd(f_i(x),G(x))=1$, $\forall i \in [n] $.
  Then, the \emph{binary generalized Goppa code} is defined by
  \begin{equation}\label{eq:gcc}
    \Gamma(\cL,G):= \left\{\c\in\F_2^n\ |\ \sum_{i=1}^{n}c_i \dfrac{f_i'(x)}{f_i(x)} = 0 \Mod G(x)\right\},
  \end{equation}
  where $f_i'(x)$ is the formal derivative of $f_i(x)$. We call $G(x)$
  the \emph{Goppa polynomial} and $\cL$ the set of \emph{code locators}\footnote{In the original definition~\cite{bezzateev1997gGC}, the code locators are defined by ${f'_i(x)}/{f_i(x)}$. For ease of notation, we define the code locators by $f_i(x)$ here.}.
\end{definition}
In the following theorem, we derive a parity-check matrix for generalized Goppa codes with code locators of arbitrary degree.
\begin{theorem}[Parity-Check Matrix]\label{thm:parityCheckMatrix}
Given a binary generalized Goppa code $\Gamma(\cL,G)$ as in~\cref{def:bgGC}, where the code locators in $\cL$ are $\Fq$-irreducible polynomials
  \begin{align*}
    f_i(x)=\prod_{j=0}^{l_i-1}\left(x-\power{\gamma_i}{j}\right),\  \forall i \in [n]
  \end{align*}
of degree $l_i$, where $\power{\gamma_i}{j}\in\mathbb{F}_{q^{l_i}}$
are the roots of $f_i(x)$. Let $r=\deg G(x)$ and $n=|\cL|$.
A parity-check matrix $\H$ of $\Gamma(\cL,G)$ such that $\H \c^\top =\0,\ \forall \c\in \Gamma(\cL,G)$ is
\begin{align}\label{eq:PCM}
    \H=[
      \h_1^\top \  \h_2^\top \  \cdots \  \h_n^\top
                                       ]\in\Fq^{r \times n}
\end{align}
with $\h_i = (h_{i,1} \ h_{i,2} \ \cdots \ h_{i,r} )$, where
\begin{equation*}
h_{i,j}  = \displaystyle \sum\limits_{\iota=0}^{l_i-1}\dfrac{\gamma^{(j-1)q^{\iota}}_{i}}{G\left(\power{\gamma_i}{\iota}\right)}, \   \forall  i \in [n],\ j \in [r].
\end{equation*}
\end{theorem}
\begin{IEEEproof}
  From~\cref{def:bgGC}, requiring~$\gcd(f_i(x),G(x))=1$ implies that the roots of $f_i(x)$ are not roots of $G(x)$, i.e.,
  $G(\power{\gamma_i}{j})\neq 0, \forall j=0,\dots, l_i-1, i \in[n]$.
  The inverse of a polynomial $f_i(x)$ can be found by the \emph{extended Euclidean (EEA) Algorithm}~\cite[Sec.~6.4]{roth2006}.
  Denote the Goppa polynomial by
  $$G(x) = G_0 + G_1 x + \dots + G_r x^r$$ with $G_r \neq 0$.
  Using the EEA, we obtain
  \begin{equation}
    \begin{split}
  	&\frac{f'_i(x)}{f_i(x)}\Mod G(x)=\left(\prod_{j=0}^{l_i-1}G\left(\power{\gamma_i}{j}\right)\right)^{-1}\cdot \\
  	&\sum_{t=0}^{r-1}x^t\left(\sum_{k=t+1}^r G_k\left(\sum_{j=0}^{l_i-1}\gamma_i^{(k-1-t)q^{j}}\prod_{\substack{\xi=0,\\ \xi\neq j}}^{l_i-1}G\left(\power{\gamma_i}{\xi}\right)\right)\right)
    \end{split} \label{eq:eeas}
  \end{equation}
  Plugging~\eqref{eq:eeas} into~\eqref{eq:gcc} and equating the coefficients of $x^t,\forall t \in[0,r-1]$ to zero, it can be verified that $\c\in\Gamma(\cL,G)$ if and only if $\G\H\cdot \c^{\top}=\0$, where
  \begin{align*}
  \G=\begin{bmatrix}
  G_r & 0 & \dots & 0\\
  G_{r-1} & G_r  & \dots & 0\\
  \vdots & \vdots & \ddots  & \vdots \\
  G_1 & G_2  & \dots & G_r
  \end{bmatrix}.
  \end{align*}
  Therefore, $\widetilde{\H}=\G\H$ is a parity-check matrix of $\Gamma(\cL,G)$. Since $\G$ is invertible, $\widetilde{\H}\cdot\c^{\top}=\mathbf{0}\Longleftrightarrow \H\cdot \c^{\top}=\mathbf{0}$, which proves the statement.
\end{IEEEproof}
A binary parity-check matrix $\H^{\mathrm{bin}}\in\F_2^{rm\times n}$ of $\Gamma(\cL,G)$ can be obtained by replacing every entry in $\H$ from~\eqref{eq:PCM} with a length-$m$ column vector representation over $\F_2$ according to some fixed basis of $\Fq$ over $\F_2$.
\begin{theorem}[Dimension, Minimum Distance]\label{thm:distanceBinary}
  Given a binary generalized Goppa code $\Gamma(\cL,G)$ as in~\cref{def:bgGC},
  the dimension is
  \begin{align}
    k(\Gamma)=n-\rank(\H^{\mathrm{bin}})\geqslant n-rm,
  \end{align}
  where $\H^{\mathrm{bin}}\in\F_2^{rm\times n}$ is the $\F_2$-representation of $\H\in\F_{2^m}^{r\times n}$ from~\cref{thm:parityCheckMatrix}.
  The minimum Hamming distance is
  \begin{equation*}
    d(\Gamma) \geqslant d_{\mathrm{g}} := \frac{r+1}{l},
  \end{equation*} where
  $l=\max_{f(x)\in\cL}\deg f(x)$.
\end{theorem}
\begin{IEEEproof}
It can be readily seen that $\H\c^{\top}=\0 \Longleftrightarrow \H^{\mathrm{bin}}\c^{\top}=\mathbf{0},\forall \c \in\Gamma(\cL,G)$. The dimension follows from the size of the parity-check matrix.
To prove the minimum Hamming distance, consider a codeword $\c\in\Gamma$.
Define
\begin{equation*}
  F_{\c}(x) :=\prod_{i \,\in \,\supp(\c)} f_i(x),
\end{equation*}
where its formal derivative is denoted as
\begin{equation*}
  F_{\c}'(x) :=\sum_{i \,\in \, \supp(\c)} f_i'(x)\prod_{\substack{j \, \in \, \supp(\c)\\j\neq i}}f_j(x).
\end{equation*}
Furthermore, let
\begin{equation}
R_{\c}(x) :=\sum_{i \, \in \, \supp(\c)} \frac{f'_i(x)}{f_i(x)}=\frac{F'_{\c}(x)}{F_{\c}(x)},\label{eq:Rx}
\end{equation}
where $f_i'(x)$ is the formal derivative of $f_i(x)$.
Since all $f_i(x)$ have distinct roots, $\gcd(F'_{\c}(x),F_{\c}(x))=1$ and
since $\gcd(f_i(x),G(x))=1,\ \forall i\in[n]$, $\gcd(F_{\c}(x),G(x))=1$.
Then from~\eqref{eq:Rx},
  \begin{align*}
    R_{\c}(x)= 0 \mod G(x)\quad \iff\quad G(x) | F'_{\c}(x).
  \end{align*}
   Note that $F'_{\c}(x)$ is the formal derivative of $F_{\c}(x)$. Since we are working over a field of characteristic $2$, $F'_{\c}(x)$ only has even powers and is a perfect square. Let $\bar{G}(x)$ be the lowest-degree perfect square which is divisible by $G(x)$, then
  \begin{align*}
    G(x)|F'_{\c}(x) \quad \iff\quad \bar{G}(x)|F'_{\c}(x). \nonumber
  \end{align*}
  Thus,
  \begin{align}
    \c\in \Gamma \quad &\iff \quad R_{\c}(x)= 0 \mod G(x)\nonumber \\
    &\iff\quad \bar{G}(x)|F'_{\c}(x).\label{eq:GbarDividesFprime}
  \end{align}
Denote $l_i=\deg f_i(x)$, then $\deg F_{\c}(x)=\sum_{i\in\supp(\c)} l_i$ and
  \begin{align}\label{eq:degFprime}
	\deg F'_{\c}(x) \leqslant \deg F_{\c}(x)-1=\sum_{i \,\in \,\supp(\c)} l_i -1.
  \end{align}
  Consider a vector \new{$\v_m$} whose support $\supp(\v_m)$ concentrates on the locators of the highest degree $l= \max_{i} l_i$, then
  \begin{align}\label{eq:minWeightDeg}
    \deg F'_{\v_m}(x) 
    &\leqslant \wt(\v_m)\cdot l-1.
  \end{align}
  \new{Note that $\v_m$ is not necessarily a codeword.}
  Let $\deg F'_{\v_m}(x)\overset{!}{\geqslant}\deg \bar{G}(x)$. We have $\wt(\v_{m})\geqslant (\deg \bar{G}(x)+1)/{l}$.
  To have~\eqref{eq:GbarDividesFprime} fulfilled, we require
  \begin{align}\label{eq:minDegFc}
    \deg F'_{\c}(x)&\geqslant \deg \bar{G}(x) \quad \forall \ \c\in\Gamma.
  \end{align}
  Note that for any $\c$ with $\wt(\c)< \wt(\v_m)$, $\deg F'_{\c}(x)<\deg F'_{\v_m}(x)$, i.e., we cannot find a codeword $\c$ with $\wt(\c)<\wt(\v_m)$ such that $\deg F'_{\c}(x)\geqslant \deg F'_{\v_m}(x)$.
  Therefore, to fulfill~\eqref{eq:minDegFc},
  \begin{align*}
    d(\Gamma)&=\min_{\c\in\Gamma}\wt(\c)\geqslant \wt(\v_m)\\
             &\geqslant \frac{\deg \bar{G}(x)+1}{l}\geqslant \frac{\deg {G}(x)+1}{l}=d_{\mathrm{g}}. \tag*{\IEEEQEDhere}
  \end{align*}
\end{IEEEproof}
Classical Goppa codes with a Goppa polynomial which has only distinct roots are known as \emph{separable} Goppa codes~\cite[Ch.~12]{macwilliams1977theory}.
In this paper we inherit this name and call the GGCs with a Goppa polynomial which has only distinct roots as \emph{separable generalized Goppa codes}.
\begin{corollary}\label{col:distanceWild}
  Given a Goppa polynomial $G(x)$ whose roots are all distinct, the binary \emph{separable} generalized Goppa code $\Gamma(\cL,G)$ is the same code as $\Gamma(\cL,G^2)$ and the minimum distance is
  \begin{equation}\label{eq:d_gGC_separable}
    d(\Gamma) \geqslant d_{\mathrm{sep}} :=\frac{2r+1}{l}.
  \end{equation}
\end{corollary}
\begin{IEEEproof}
Since all roots of $G(x)$ are distinct, $\bar{G}(x)=G(x)^2$ in the proof of~\cref{thm:distanceBinary}. The statement follows herein.
\end{IEEEproof}
\new{The following corollary shows that}
with \new{even-degree} code locators, the lower bound on the minimum distance is \new{increased by a difference of ${1}/{l}$ compared to \eqref{eq:d_gGC_separable}}.
  \begin{corollary}
    Given a code locator set $\cL$ of \emph{even-degree} polynomials, the minimum distance of a binary separable generalized Goppa code $\Gamma(\cL,G)$ is
    \begin{align*}
      d(\Gamma) \geqslant d_{\mathrm{even}} :=\frac{2r+2}{l}.
    \end{align*}
  \end{corollary}
  \begin{IEEEproof}
    Since $\deg f_i(x)$ is even for all $f_i(x)\in\cL$, $\deg F_{\c}(x)$ is even. Then $\deg F'_{\c}(x)\leqslant \sum_{i\in\supp(\c)}l_i-2$ in~\eqref{eq:degFprime} and $\deg F'_{\v_m}(x)\leq \wt(\v_m)\cdot l-2$ in~\eqref{eq:minWeightDeg} since we work over a field of characteristic $2$.
    Together with the separable property from~\cref{col:distanceWild}, the statement follows from the rest of the proof of~\cref{thm:distanceBinary}.
  \end{IEEEproof}
Compared to classical Goppa codes, the code length $n$ of the generalized Goppa codes is not limited by the field size $q=2^m$, but by the number of irreducible polynomials in $\F_{2^m}[x]$. The result in the following theorem was stated in~\cite{noskov_one_2020}. As a completion to~\cref{thm:parityCheckMatrix} and~\cref{thm:distanceBinary} for the properties of binary generalized Goppa codes, we include it here.
\begin{theorem}[Code Length~{\cite{noskov_one_2020}}]\label{thm:generalizedLength}
Let $q=2^m$ for some integer~$m$. Given a generalized Goppa code $\Gamma(\cL,G)$. Denote $l=\max_{f(x)\in\cL}\deg f(x)$.
The length of $\Gamma(\cL,G)$ is limited by
\begin{equation}
  n(\Gamma) \leqslant \sum_{t=1}^{l}\mathcal{I}_{q}(t),
  \label{eq:generalized_length}
\end{equation}
where $\mathcal{I}_{q}(t)$ is the number of irreducible polynomials of degree~$t$ in the polynomial ring $\mathbb{F}_{q}[x]$
(see~\cref{lem:numIrrePoly}).
\end{theorem}

\section{Decoding of Generalized Goppa Codes}\label{sec:UniDec}
In this section, we \new{present an explicit decoding algorithm for GGCs, where the decoding principle has been mentioned in~\cite{noskov_one_2020}. This syndrome-based decoding algorithm is also a basis of the joint decoder for interleaved GGCs, which we present in~\cref{sec:IntDec}. Moreover, we show that the unique decoding radius for GGCs is $\floor{\frac{d}{2}}$, which is different from the usual form $\floor{\frac{d-1}{2}}$ for other codes by such decoding algorithm. }
\begin{definition}\label{def:GoppaSyndrome}
  Consider a binary generalized Goppa code $\Gamma(\cL,G)$ and an error vector $\e\in\F_2^n$ where $n=|\cL|$. Let $\cE=\supp(\e)$.
  Define the \emph{syndrome polynomial}
  \begin{equation}\label{eq:goppasyndromedef}
    s(x) := \sum\limits_{i\in\cE}e_i\frac{f'_i(x)}{f_i(x)}\mod G(x),
  \end{equation}
  the \emph{error locator polynomial (ELP)}
  \begin{equation}\label{eq:goppaELP}
    \Lambda(x) := \prod\limits_{i\in\set{E}}f_i(x),
  \end{equation}
  and the \emph{error evaluator polynomial (EEP)}
  \begin{align}\label{eq:goppaEEP}
    \Omega(x) := \sum\limits_{i\in\set{E}}e_if'_i(x) \prod\limits_{j\in \set{E}\backslash\{i\}}f_j(x).
  \end{align}
\end{definition}
Assume transmitting a codeword $\c\in\Gamma(\cL,G)$ and receiving a vector $\r=\c+\e\in\F_2^n$. The syndrome polynomial can be calculated from the received word $\r$ by
  \begin{equation}\label{eq:goppasyndromecal}
    s(x) = \sum\limits_{i=1}^{n}r_i\frac{f'_i(x)}{f_i(x)}\mod G(x).
  \end{equation}
  Denote $s(x) = \sum_{i=1}^{r} s_i x^{r-i}$ where $\s=(s_1,\dots,s_{r})=\r\widetilde{\H}^{\top}$.

  We present a syndrome-based decoder for $\Gamma(\cL,G)$ in~\cref{algo:SyndromeDecoder}.
  The main step of decoding is to determine $\Lambda(x)$ and $\Omega(x)$ given $s(x)$. In the following lemma we set up a key equation for decoding generalized Goppa codes.
  \begin{lemma}[Key Equation]\label{lem:GoppaKeyEq}
    Consider a binary generalized Goppa code $\Gamma(\cL,G)$. 
    Assume an error $\e$ of weight $t$ occurs.
    Then, the following equations hold, which are called the \emph{key equation} for decoding $\Gamma(\cL,G)$: 
    \begin{align}
      \Omega (x) & = \Lambda(x) s(x) \mod G(x) \label{eq:goppakeyeq}\\
      \gcd(\Lambda(x),\Omega(x)) & = 1 \label{eq:goppakeygcd}\\
      \deg\Omega(x) & < \deg \Lambda(x)\leqslant t\cdot l \label{eq:goppakeydeg}
    \end{align}
    where $l=\max_{f(x)\in\cL}\deg f(x)$.
\end{lemma}
\begin{IEEEproof}
  Denote $\cE=\supp(\e)$ and $t=|\cE|$.
  Eq.~\eqref{eq:goppakeyeq} follows from~\eqref{eq:goppasyndromedef}, \eqref{eq:goppaELP}, and \eqref{eq:goppaEEP} since
  \begin{align*}
    s(x) = \frac{\sum\limits_{i\in\set{E}}e_if'_i(x) \prod\limits_{j\in \set{E}\backslash\{i\}}f_j(x)}{\prod\limits_{i\in\set{E}}f_i(x)}=\frac{\Omega(x)}{\Lambda(x)}\mod G(x).
  \end{align*}
  Eq.~\eqref{eq:goppakeygcd} holds since all $f_i(x)$ have distinct roots.
  From the definitions of ELP in~\eqref{eq:goppaELP}, $\deg \Lambda(x)=\sum_{i\in\cE} \deg f_i(x)\leqslant t\cdot \max_{i\in\cE}\deg f_i(x)\leqslant t\cdot l$. From~\eqref{eq:goppaEEP}, $\deg \Omega(x)=\deg \Lambda'(x)<\deg \Lambda(x)$. The degree constraints in~\eqref{eq:goppakeydeg} follow herein.
\end{IEEEproof}
\begin{theorem}[Unique Decoding Radius]\label{thm:t_uni}
  Given a binary \emph{separable} generalized Goppa code $\Gamma(\cL,G)$ with $d(\Gamma)\geqslant d_{\mathrm{sep}}$, \cref{algo:SyndromeDecoder} can uniquely decode any error $\e$ of weight
  \begin{align*}
    t \leqslant t_{\mathrm{sep}}:=\floor{\frac{r}{l}}= \floor{\frac{d_{\mathrm{sep}}}{2}},
  \end{align*}
  where $r= \deg G(x)$ and $l=\max_{f(x)\in\cL}\deg f(x)$.
\end{theorem}
\begin{IEEEproof}
  It follows from~\cite[Proposition 6.3, 6.4]{roth2006} that~\cref{step:calELPEEP} of \cref{algo:SyndromeDecoder}
  will find a unique solution of the pair $(\lambda(x),\omega(x))$ such that $\Lambda(x)=c\cdot \lambda(x), \Omega(x)=c\cdot\omega(x)$ for some constant $c$, if $\deg \omega(x)<\deg \lambda(x)\leqslant {\deg (G(x)})/{2}$.
  At~\cref{step:chienSearch} of \cref{algo:SyndromeDecoder} we search for the roots of $\lambda(x)$. They are also roots of $\Lambda(x)$ if $\deg \Lambda(x)=\deg\lambda(x)\leqslant {\deg (G(x))}/{2}$. Namely, the error locations can be uniquely determined if $\deg\Lambda(x)\leqslant \deg (G(x))/{2}$.

  Since the separable generalized Goppa code $\Gamma(\cL,G)$ is the same code as $\Gamma(\cL,G^2)$ according to~\cref{col:distanceWild}, we can apply~\cref{algo:SyndromeDecoder} on $\Gamma(\cL,G^2)$ to decode $\Gamma(\cL,G)$. Then, the degree constraint for uniquely decoding $\Lambda(x)$ becomes ${\deg (G(x)^2)}/{2}$.
  Thus,
  \begin{align*}
    \deg \Lambda(x)\leqslant t\cdot l \overset{!}{\leqslant} \frac{\deg (G(x)^2)}{2} = r.
  \end{align*}
  It holds that ${r}/{l}<{r}/{l}+{1}/{(2l)}={d_{\mathrm{sep}}}/{2}$.
  In particular, $\floor{{r}/{l}}<\floor{{r}/{l}+{1}/{(2l)}}$ only if $2l|(2r+1)$, which is impossible for positive integers $r$ and $l$.
  Therefore the equality holds.
\end{IEEEproof}
\begin{algorithm}[htb]
  \caption{Syndrome-based Decoding Algorithm}\label{algo:SyndromeDecoder}
  \SetAlgoLined
  \DontPrintSemicolon
  \KwIn{Code $\Gamma(\cL,G)$, received word $\r \in \F_2^n$}
  \vspace{.1cm}
  Calculate $s(x)$ by~\eqref{eq:goppasyndromecal}\;
  $\omega(x),\_,\lambda(x)$ $\gets$ \texttt{EEA}$(G(x),s(x))$ \label{step:calELPEEP}
  with the stopping condition that $\deg\omega(x)<\deg\lambda(x)\leqslant\deg G(x)/{2}$ \tcp*[r]{See~\cite[Sec.~6.4]{roth2006} for $\mathtt{EEA}$}
  $\cE\gets \{i:\lambda(\gamma_i)=0\}\ ^{*}$\tcp*{$\gamma_i$ is a root of $f_i(x)$}\label{step:chienSearch}
  $ \e\gets\0; e_i  \gets 1, \; \forall i\in\cE$\;
  \vspace{.1cm}
  \KwOut{$\hat{\c} \gets \r - \e$}
  \SetAlgoLined
  \algorithmfootnote{$^{*}$Verifying $\lambda(\gamma_i)=0$ can be done by applying Chien Search~\cite{DecRS1964Chien} in each splitting field $\F_{q^{l_i}}$ if there is an $f_i(x)\in\cL$ of degree $l_i$ and we only need to do this evaluation at one of the roots of $f_i(x)$.  \vspace{-1em}}
\end{algorithm}
\section{Joint Decoding of Interleaved Generalized Goppa Codes}\label{sec:IntDec}
Interleaved codes are known to be able to decode beyond the unique decoding radius~\cite{metzner1990general,bleichenbacher2003decoding,justesen2004decoding,schmidt2005interleaved,holzbaur2019decoding}, especially in appearance of \emph{burst errors}. Burst errors can be modelled as an error matrix $\E$ that only has a few non-zero columns. We denote by $\supp(\E)$ the indices of the non-zero columns of $\E$.
\new{We present the explicit decoder and derive the new maximum decoding radius for GGCs, which is different from the general form of that for interleaved Reed-Solomon codes~\cite{schmidt2005interleaved} or interleaved classical Goppa codes~\cite{holzbaur2019decoding}. }
\begin{definition}[Interleaved Generalized Goppa Codes]\label{def:bgGCInt}
	Let $w$ be the interleaving order. Given a generalized Goppa code $\Gamma(\mathcal{L},G)$,
	a $w$-\emph{interleaved generalized Goppa code} is denoted by $w$-$\mathcal{I}\Gamma(\mathcal{L},G)$ and defined by
	\begin{equation*}
	w\text{-}\mathcal{I}\Gamma(\mathcal{L},G):=
	\left\lbrace
	\begin{pmatrix}
	\c^{(1)}\\
	\vdots\\
	\c^{(w)}\\
	\end{pmatrix}, \ \forall \c^{(i)}\in \Gamma(\cL,G), \ i\in[w]
	\right\rbrace.
	\end{equation*}
\end{definition}
Consider transmitting a codeword $\C\in w$-$\cI\Gamma(\cL,G)$ with $n=|\cL|$. An error $\E\in\F_2^{w\times n}$ with $\cE=\supp(\E)$ occurs and we receive $\R=\C+\E$.
We follow the definitions from~\cref{def:GoppaSyndrome} for the ELP $\Lambda(x)$, the syndromes $s^{(i)}(x)$ and the EEPs $\Omega^{(i)}(x)$ for $\E$ with $\cE=\supp(\E)$.

\begin{lemma}[Key Equations for Joint Decoding]\label{lem:keyEquationInt}
  The \emph{key equations} for decoding $w$-$\cI\Gamma(\cL,G)$ in occurrence of an error $\E$ with $t$ non-zero columns are:
  \begin{align*}
    \Omega^{(i)} (x) &= \Lambda(x) s^{(i)}(x) \mod G(x)\\
    \deg\Omega^{(i)}(x)&<\deg \Lambda(x) \leqslant t\cdot l
  \end{align*}
  for all $i\in [w]$, where $l=\max_{f(x)\in\cL}\deg f(x)$.
\end{lemma}
Instead of solving the key equations in~\cref{lem:keyEquationInt} for the $\Lambda(x)$ and $\Omega^{(i)}(x)$ which have specific algebraic structures, we solve the following general version of this problem:
Given $G(x), s^{(1)}(x), \dots, s^{(w)}(x)\in\Fq[x]$, find a lowest-degree polynomial $\lambda(x)$ such that there exist polynomials $\omega^{(1)}(x), \dots, \omega^{(w)}(x)\in\F_q[x]$, not all zero, satisfying
\begin{equation}\label{eq:keyEquationLinear}
  \begin{split}
    \omega^{(i)} (x) &= \lambda(x) s^{(i)}(x) \mod G(x)\\
    \deg\omega^{(i)}(x)&<\deg \lambda(x)\leqslant t\cdot l
  \end{split}
\end{equation}
for all $i\in[w]$.
This problem can be solved by the \emph{MgLFSR Algorithm}~\cite{nielsen2013generalised}, by the \emph{Feng-Tzeng Euclidean algorithm}~\cite{ZehWachter2011fast},
or by solving a \emph{linear system of equations} (LSE) for the unknown coefficients of $\lambda(x)$~\cite[Sec.~4.3.2]{liu_decoding_master}. We summarize the decoding procedure in~\cref{algo:IntDec}.

\begin{theorem}[Maximum Decoding Radius]\label{thm:tInt}
  Given a binary interleaved separable generalized Goppa code $w$-$\cI\Gamma(\cL,G)$ with $d(\Gamma)\geqslant d_{\mathrm{sep}}$, with high probability, \cref{algo:IntDec} can decode an error $\E$ with $t$ non-zero columns if
  \begin{align*}
    t\leqslant t_{\max}:=\floor{\frac{w}{w+1}\cdot\frac{2r}{l}} \leqslant
    \floor{\frac{w}{w+1}\cdot d_{\mathrm{sep}}}
  \end{align*}
   where $r= \deg G(x)$ and $l=\max_{f(x)\in\cL}\deg f(x)$.
\end{theorem}
\begin{IEEEproof}
  Note that the separable $w$-$\cI\Gamma(\cL,G)$ is the same code as $w$-$\cI\Gamma(\cL,G^2)$, therefore we can decode $w$-$\cI\Gamma(\cL,G)$ by applying~\cref{algo:IntDec} on $w$-$\cI\Gamma(\cL,G^2)$.
  By setting up the LSE for~\eqref{eq:keyEquationLinear} according to~\cite[Sec.~4.3.2]{liu_decoding_master},
  we can get
  \begin{align}\label{eq:numOfEquations}
    \deg (G(x)^2)-\deg \omega^{(i)}-1= \deg (G(x)^2)-\deg \lambda(x)
  \end{align}
  equations for $\deg\lambda(x)$ unknowns (i.e., coefficients of $\lambda(x)$) from each congruence. The unknowns are the same for every congruence.
  In total we have at most $w(\deg (G(x)^2)-\deg \lambda(x))$ equations for $\deg\lambda(x)$ unknowns. To have a unique solution, the number of unknowns should not be more than the number of equations, i.e.,
  \begin{align}
  \deg\lambda(x)&\leqslant w(\deg (G(x)^2)-\deg \lambda(x)),\nonumber\\
    \deg\lambda(x)&\leqslant \frac{w}{w+1}\deg (G(x)^2).\label{eq:degLimitProof}
  \end{align}
  Suppose $\Lambda(x)=c\cdot \lambda(x)$, $\Omega^{(i)}(x)=c\cdot \omega^{(i)}(x),\forall i \in[w]$. Then, we can get a unique solution for $\Lambda(x), \Omega^{(i)}(x)$ by~\cref{algo:IntDec} if the solution for $\lambda(x)$ is unique, i.e., if~\eqref{eq:degLimitProof} is fulfilled. The second inequality in the statement holds by plugging in $d_{\mathrm{sep}}$ from~\eqref{eq:d_gGC_separable}.
\end{IEEEproof}

\begin{corollary}\label{col:tInt_even}
Given a binary interleaved separable generalized Goppa code $w$-$\cI\Gamma(\cL,G)$ with all code locators of even-degree, with high probability, \cref{algo:IntDec} can decode an error $\E$ with $t$ non-zero columns if
 \begin{align*}
    t\leqslant t^{(\mathrm{even})}_{\max}:= \floor{\frac{w}{w+1}\cdot\frac{2r+1}{l}} \leqslant \floor{\frac{w}{w+1}\cdot d_{\mathrm{even}}},
  \end{align*}
where $r= \deg G(x)$ and $l=\max_{f(x)\in\cL}\deg f(x)$.
\end{corollary}
\begin{IEEEproof}
For only even-degree code locators, $\deg\Omega^{(i)}(x)\leqslant \deg \Lambda(x)-2$ since we work in a field of characteristic $2$. Therefore, when setting up the LSE, instead of~\eqref{eq:numOfEquations}, we will have $\deg G(x)-\deg \lambda(x)+1$ equations from each congruence. The rest of the proof remains the same as for~\cref{thm:tInt}.
 \end{IEEEproof}
 The maximum decoding radius $t^{(\mathrm{even})}_{\max}$ for interleaved separable GGCs with even-degree code locators can be increased by $1$ upon $t_{\max}$ in~\cref{thm:tInt} if and only if we choose the interleaving order $w$ such that $(w+1)|(2r+1)$ and $l|w$.
\begin{algorithm}[htb!]
  \caption{Decoding Algorithm for $\cI\Gamma(\cL,G)$}\label{algo:IntDec}
  \SetAlgoLined
  \DontPrintSemicolon
  \KwIn{$w$-$\cI\Gamma(\cL,G)$, received word $\R \in \F_2^{w\times n}$}
    \vspace{.1cm}
  Calculate $w$ syndromes $s^{(i)}(x), \; \forall i \in [w]$ by~\eqref{eq:goppasyndromecal}\;
  Solve~\eqref{eq:keyEquationLinear} for $\lambda(x)$ by solving LSE~\cite{liu_decoding_master}, MgLFSR~\cite{nielsen2013generalised} or Feng-Tzeng EEA~\cite{ZehWachter2011fast}\;
  \textbf{If} $\lambda(x)$ is not unique {\textbf{Return} \texttt{decoding failure}}\;
  $\cE\gets \{i:\lambda(\gamma_i)=0\}\ ^{*}$\tcp*{$\gamma_i$ is a root of $f_i(x)$}
  Calculate $\omega^{(i)}(x)=\lambda(x)s^{(i)}(x)\Mod G(x),\ \forall i\in[w]$\;
  $\E\gets \0$;\quad denote by ${e}^{(i)}_j$ the $(i,j)$-entry of $\E$\;
  \lForEach{$i\in[w],j\in\cE$}{${e}^{(i)}_j={\omega^{(i)}(\gamma_j)}/{\lambda'(\gamma_j)}\ ^{**}$}
  \tcp*[r]{$\gamma_j$ is a root of $f_j(x)$}
    \vspace{.1cm}
  \KwOut{$\widehat{\C}=\R-\E$ or \texttt{decoding failure}}
  \SetAlgoLined
  \algorithmfootnote{$^*$Apply Chien Search~\cite{DecRS1964Chien} for fast implementation.
    See also in~\cref{algo:SyndromeDecoder}.\\$^{**}$This follows from \emph{Forney's Algorithm}~\cite{Forney1965}.}
\end{algorithm}
\begin{remark}
  \cref{algo:IntDec} may output \texttt{decoding failure} if the number of errors $t>t_\mathrm{sep}$ in~\cref{thm:t_uni}.
  The failure results from the linear dependency of equations in the LSE. An upper bound on the failure probability of decoding interleaved alternant codes has been recently derived in~\cite{holzbaur2020IABound}, which holds for decoding interleaved GGCs with code locators of degree one. 
\end{remark}
\section{Code Parameters} \label{sec-McEliece}
\label{sec:crypto}

In Table \ref{tab:codeparameters}, we show some examples of code parameters ($k \geqslant$, $m$, $l$, $r$, $d_{\mathrm{sep}}$) of binary separable Goppa codes and binary separable generalized Goppa codes $\Gamma(\cL,G(x))$\new{, denoted by GC and GGC-$l$ respectively,} for several values of length~$n$.

For \new{GGCs and} a fixed code length $n$, the degree~$m$ of the extension field can be reduced according to \eqref{eq:generalized_length} by increasing the maximum degree~$l$ of the code locators in $\cL$.
By additionally fixing the degree~$r$ of the Goppa polynomial, the lower bound on the minimum distance $d_{\mathrm{sep}}$ is reduced by the factor of~$l$, according to~\cref{col:distanceWild}.
Keeping instead $d_{\mathrm{sep}}$ fixed, the degree $r$ must be increased to $r=\ceil{(l\cdot d_{\mathrm{sep}}-1)/{2}}$.
The lower bound on the dimension $k$ is calculated by $n-mr$ and is therefore smaller for a higher degree of~$r$.
\new{The specialty of GGCs is that the code length $n$ can be greater than the size of the extension field.}
\vspace{-.4cm}

\begin{table}[htb]
	\centering
	\caption{Code parameters for binary separable GGCs.}
	\label{tab:codeparameters}
	\begin{tabular}{l|l|clllrr}
		Code &${n}$ & ${k}\geqslant$ & ${m}$ & ${l}$ & ${r}$ &  ${d_{\mathrm{sep}}}$ & \textit{$\mathit{|\mathsf{pk}|}$~[bytes]} \\
		\hline
		GC &3488 &   2720  & 12 & 1 & 64 & 129 & \textit{261 120} \\
		GGC-2 &3488 & 3040  & 7 & 2 & 64 & 64 & \textit{170 240} \\
        GGC-2 &3488 & 2585 & 7 & 2 & 129 & 129 & \textit{291 782} \\

		\hline

		GC &6960 &5413 & 13 & 1& 119 & 239 & \textit{1 047 319}  \\
		GGC-2 &6960 & 6127 & 7 & 2 & 119 & 119 & \textit{637 974}  \\
		GGC-3 &6960 & 5170 & 5 & 3 & 358 & 239  & \textit{1 156 788}  \\

		\hline
		GC &8192 & 6528 & 13 & 1 & 128 & 257 & \textit{1 357 824}  \\ 
		GGC-2 &8192 &7296 & 7 & 2 & 128 & 128 & \textit{817 152}  \\

		GGC-8 &8192 &6528 & 2 & 8 & 832 & 208 & \textit{1 357 824}\\

	\end{tabular}
\end{table}

Table \ref{tab:codeparameters} also shows the corresponding public key size of \textit{Classic McEliece} \cite{ClassicMcElieceNIST}, which is the Niederreiter's dual version of the original McEliece cryptosystem and currently a finalist of the NIST competition for post-quantum key encapsulation mechanisms \cite{computer_security_division_round_2017}.
The cryptosystem is efficient in encoding and decoding, but \new{it has a large public key size, which is a drawback in computation time and storage space.}
The public key~$\mathbf{T}$ is determined by the systematic form of $\H_\mathrm{bin}=(\mathbf{I}_{n-k} \; | \; \mathbf{\mathbf{T}})$ and has size $|\mathsf{pk}|=(nmr-m^2r^2)/8$ bytes.

\new{The complexity of all computations including the construction of a parity-check matrix (public key) can be improved by reducing the field size with GGCs. The cost is a larger public key size or a smaller security level, based on the \emph{Information Set Decoding} (ISD) attack by Lee and Brickell \cite{lee_observation_1988}, whose work factor depends on $n,k,d$.}

\bibliographystyle{IEEEtran}
\bibliography{main}

\end{document}